\documentclass[aps,prl,showpacs,twocolumn,superscriptaddress,linenumbers,floatfix]{revtex4}

\usepackage{graphicx}  
\pdfoutput=1
\usepackage{amsmath,bm}   

\usepackage{dcolumn}

\begin{document}


\title{Determination of the nitrogen vacancy as a shallow compensating center in GaN doped with divalent metals}

\author{J. Buckeridge} 
\email{j.buckeridge@ucl.ac.uk}
\affiliation{University College London, Kathleen Lonsdale Materials
  Chemistry, Department of Chemistry, 20 Gordon Street, London WC1H
  0AJ, United Kingdom}
\author{C. R. A. Catlow}
\affiliation{University College London, Kathleen Lonsdale Materials
  Chemistry, Department of Chemistry, 20 Gordon Street, London WC1H
  0AJ, United Kingdom}
\author{D. O. Scanlon}
\affiliation{University College London, Kathleen Lonsdale Materials
  Chemistry, Department of Chemistry, 20 Gordon Street, London WC1H
  0AJ, United Kingdom}
\affiliation{Diamond Light Source Ltd., Diamond House, Harwell Science
  and Innovation Campus, Didcot, Oxfordshire OX11 0DE, United Kingdom}
\author{T. W. Keal}
\affiliation{Scientific Computing Department, STFC, Daresbury
  Laboratory, Daresbury, Warrington, WA4 4AD, United Kingdom}
\author{P. Sherwood}
\affiliation{Scientific Computing Department, STFC, Daresbury
  Laboratory, Daresbury, Warrington, WA4 4AD, United Kingdom}
\author{M. Miskufova}
\affiliation{University College London, Kathleen Lonsdale Materials
  Chemistry, Department of Chemistry, 20 Gordon Street, London WC1H
  0AJ, United Kingdom}
\author{A. Walsh}
\affiliation{Centre for Sustainable Chemical Technologies and
  Department of Chemistry, University of Bath, Claverton Down, Bath
  BA2 7AY, United Kingdom}
\author{S. M. Woodley}
\affiliation{University College London, Kathleen Lonsdale Materials
  Chemistry, Department of Chemistry, 20 Gordon Street, London WC1H
  0AJ, United Kingdom}
\author{A. A. Sokol}
\email{a.sokol@ucl.ac.uk}
\affiliation{University College London, Kathleen Lonsdale Materials
  Chemistry, Department of Chemistry, 20 Gordon Street, London WC1H
  0AJ, United Kingdom}

\begin{abstract}
We report accurate energetics of defects introduced in GaN on doping
with divalent metals, focussing on the technologically important case
of Mg doping, using a model which takes into consideration both the
effect of hole localisation and dipolar polarisation of the host
material, and includes a well-defined reference level. Defect
formation and ionisation energies show that divalent dopants are
counterbalanced in GaN by nitrogen vacancies and not by holes, which
explains both the difficulty in achieving $p$-type conductivity in GaN
and the associated major spectroscopic features, including the
ubiquitous 3.46 eV photoluminescence line, a characteristic of all
lightly divalent-metal-doped GaN materials that has also been shown to
occur in pure GaN samples. Our results give a comprehensive
explanation for the observed behaviour of GaN doped with low
concentrations of divalent metals in good agreement with relevant
experiment.

\end{abstract}
\pacs{61.72.J-, 71.55.Eq, 78.55.Cr}

\maketitle

Gallium nitride, a wide-gap semiconductor, is an important material
used for a range of applications, e.g. solid state lighting, high
power microelectronics, is an essential component of commercial blue
light emitting diodes, which requires both $n$- and $p$-type
conducting layers~\cite{gan_led_history_ieee_nakamura2013}. While it
has been intensively researched for several decades, many puzzling
features still remain. For instance, GaN is natively $n$-type,
indicating the presence of shallow donors, but whether these centres
are native point defects (e.g. nitrogen vacancies) or unwanted
impurities incorporated during growth (such as hydrogen, oxygen or
silicon) remains a source of
controversy~\cite{gan_aln_inn_review_jvacscitechb_strite1992,
  handbook_nitrides_morkoc}.

Industrially, the production of $p$-type GaN is still the main
bottleneck. The only successful approach has been to dope with Mg but
there are many associated problems: large concentrations of Mg close
to the solubility limit, as well as processing such as thermal
annealing or electron irradiation, are required to achieve significant
$p$-type activation; high dislocation densities limit hole mobilities;
there is a residual $n$-type concentration that must be overcome; and
self-compensation may limit
activation~\cite{iii-nitrides_leds_seong2013,
  gan_expt_n_p_type_si_jmaterres_chuah2007}.

At low temperature ($T$), the band gap of GaN is 3.503
eV~\cite{gan_expt_bandgap_prb_monemar1974}. Experimental techniques
such as photoconductance, photoluminescence (PL), and deep level
transient spectroscopy indicate that the Mg$_{\text{Ga}}$ acceptor
state lies $\sim0.150 - 0.265$ eV above the valence band maximum
(VBM)~\cite{semiconductor_handbook_madelung2004}, corrobarated by
theoretical studies~\cite{GaN_review,
  gan_dft_mgdoped_2level_apl_lany2010}. With such a deep level,
however, why Mg-doping results in $p$-type conductivity is not well
understood, although theories have been proposed involving defect
complexes and hydrogen
impurities~\cite{gan_dft_vn_lum_mg-doped_apl_yan2012,
  gan_dft_mg_defect_holetrap_hybrid_prl_lyons2012,
  gan_dft_expt_mgdoped_mghcomplex_prl_lee2014}, with some experimental
support~\cite{gan_expt_mgdoped_epr_prb_glaser2002,
  gan_expt_mgdoped_epr_h_jap_zvanut2011}. Similar acceptor levels are
found for other divalent dopants including Zn, Cd, Be and
Sr~\cite{gan_expt_pl_mgdoped_jap_reschikov2005}, but the lack of
associated $p$-type activation, in contrast to Mg, has not been
explained.

Adding small amounts of Mg to GaN leads to characteristic peaks in the
UV range of the PL spectrum, at 3.27 eV and at 3.466 eV (measured at
$T$ = 2
K)~\cite{gan_expt_mgdoped_pl_old+new_2lines_mg_jap_monemar2014} and in
some samples a peak at $\sim3.2$
eV~\cite{gan_expt_mgdoped_pl_290mev_level_apl_smith1996,
  gan_expt_mgdoped_pl_2levels_prl_monemar2009}. The peak at 3.466 eV
is also observed in samples doped with other divalent
cations~\cite{gan_expt_pl_mgdoped_jap_reschikov2005} and is attributed
to acceptor-bound excitons
(ABEs)~\cite{gan_expt_undoped_3.466ev_acceptor_prl_stepniewski2003}.
Interestingly, there is some evidence that this peak occurs in
nominally undoped GaN
samples~\cite{gan_expt_mgdoped_pl_old+new_2lines_mg_jap_monemar2014},
indicating that it may instead relate to some compensating native
shallow defect. When the Mg content increases, the PL spectrum
changes, becoming dominated by a broad peak at $\sim2.9$ eV,
corresponding to blue luminescence (BL) and attributed to
donor-acceptor pairs
(DAPs)~\cite{gan_expt_mgdoped_pl_old+new_2lines_mg_jap_monemar2014}.
BL has also been observed in samples when $T$ is raised, rather than
the Mg concentration
([Mg])~\cite{gan_expt_mgdoped_pl_290mev_level_apl_smith1996}.  So far
theory fails to account for the characteristic UV luminescence
associated with small [Mg], but recombination processes have been
proposed to account for BL and also red
luminescence~\cite{gan_dft_mgdoped_2level_apl_lany2010,
  gan_dft_vn_lum_mg-doped_apl_yan2012,
  gan_dft_mg_defect_holetrap_hybrid_prl_lyons2012,
  gan_dft_expt_mgdoped_mghcomplex_prl_lee2014}, which is sometimes
observed in heavily-doped samples.

Recent thermal admittance spectroscopy (TAS) measurements on
epitaxially grown $n$-GaN have found that the shallow donor level lies
51 meV below the conduction band minimum
(CBM)~\cite{n_vac_optic_admittance_evwaraye_jap_2014}, agreeing with
older measurements using electron irradiation
techniques~\cite{gan_expt_eirradiate_donor_nvac_prl_look1997} (placing
the level at $64 \pm 10$ meV below the CBM). This level is
significantly deeper than those introduced when GaN is intentionally
doped $n$-type with e.g. Si or O ($\sim30$
meV)~\cite{gan_expt_pl_mgdoped_jap_reschikov2005}. Such a shallow
positive defect may act as a compensating centre for holes in
GaN, though this has not conclusively been shown to be the case.

In this Letter, we present calculated formation and ionisation
energies of nitrogen vacancies (V$_{\text{N}}$) and Mg substituting on
a Ga site (Mg$_{\text{Ga}}$) in GaN in the dilute limit, using a
multiscale embedded cluster approach. We also calculate the ionisation
energies associated with the divalent substitutionals:
Be$_{\text{Ga}}$, Zn$_{\text{Ga}}$, Cd$_{\text{Ga}}$, and
Hg$_{\text{Ga}}$ in the dilute limit. We find isolated
Mg$_{\text{Ga}}$ do not contribute to $p$-type conductivity, instead
they are compensated by V$_{\text{N}}$ formation. The same result
follows for the other divalent dopants considered. We find the
V$_{\text{N}}$ is a shallow donor at 44 meV below the CBM, accounting
for the observed 3.466 eV PL peak, and is stable in the negative
charge state, facilitating Fermi level pinning close to (and above)
the CBM and leading to native $n$-type conductivity. We determine that
the equilibrium Mg$_{\text{Ga}}$ level lies 0.307 eV above the VBM; we
also calculate related levels that depend on the hole configuration
and final spin state after excitation, that, although are slightly
less favorable, will be accessible in fast PL experiments and account
for the main peaks observed at low $T$ and [Mg]. The
observed BL is related to excitation and recombination of ionised
V$_{\text{N}}$, which will occur in greater concentrations at higher
$T$ and [Mg], in agreement with experiment. We also find
good agreement between our calculated ionisation energies for Be, Zn,
Cd, and Hg impurities and relevant PL measurements. Our results give a
simple but comprehensive explanation to the observed native $n$-type
conductivity, PL spectrum at low Mg content, and the difficulty with
$p$-type doping in GaN.

We employed the hybrid quantum mechanical/molecular mechanical (QM/MM)
embedded cluster method~\cite{
  zno_qmmm_chemsh_intjinorgchem_sokol2004} to calculate bulk and
defect energies in GaN. In this approach, a defect (possibly charged)
and its immediately surrounding region, of the order of 100 atoms, is
treated using a QM level of theory - here using density functional
theory (DFT) with a triple-zeta-plus-polarisation Gaussian basis set
(see Ref.~\cite{doping_limits_chemmater_walsh2013} for details) and a
hybrid exchange and correlation functional employing 42\% exact
exchange (BB1K)~\cite{functional_bb1k_jphyschema_zhao2004}. The QM
region is embedded within a larger cluster, typically $10000 - 20000$
atoms, which is treated at a MM level of theory, using
polarisable-shell interatomic potentials that accurately reproduce the
dielectric, elastic, and structural properties of the bulk
material~\cite{energy_review_philtransroysoca_catlow2010}. Thus one
can model accurately isolated charged defects in any
dimensionality~\cite{cds_qmmm_linear_chain_jcp_buckeridge2013}, fully
accounting for the dielectric response from the surrounding
material. As supercells are not employed, there is no need to correct
for image-charge interactions. Crucially, ionisation energies can be
determined relative to a well-defined reference level, which we term
the 'vacuum level' (although in reality it is a 'quasi-vacuum' level,
because surface effects are not included in the energetics).
Technical details, including the treatment of cluster termination and
the interface between the QM and MM regions, are discussed
elsewhere~\cite{zno_qmmm_chemsh_intjinorgchem_sokol2004,
  doping_limits_chemmater_walsh2013}. One caveat is that the defects
modelled must be localised within the QM region, which, however,
applies in the current case. This method was applied successfully to
treat defects in ZnO~\cite{zno_defects_qmmm_faraddisc_sokol2007} and
band alignment in TiO$_2$~\cite{tio2_bandalign_natmater_scanlon2013}.

\begin{figure}[ht!]
\centering
\vspace{0.2cm}
\includegraphics*[width=1.0\linewidth]{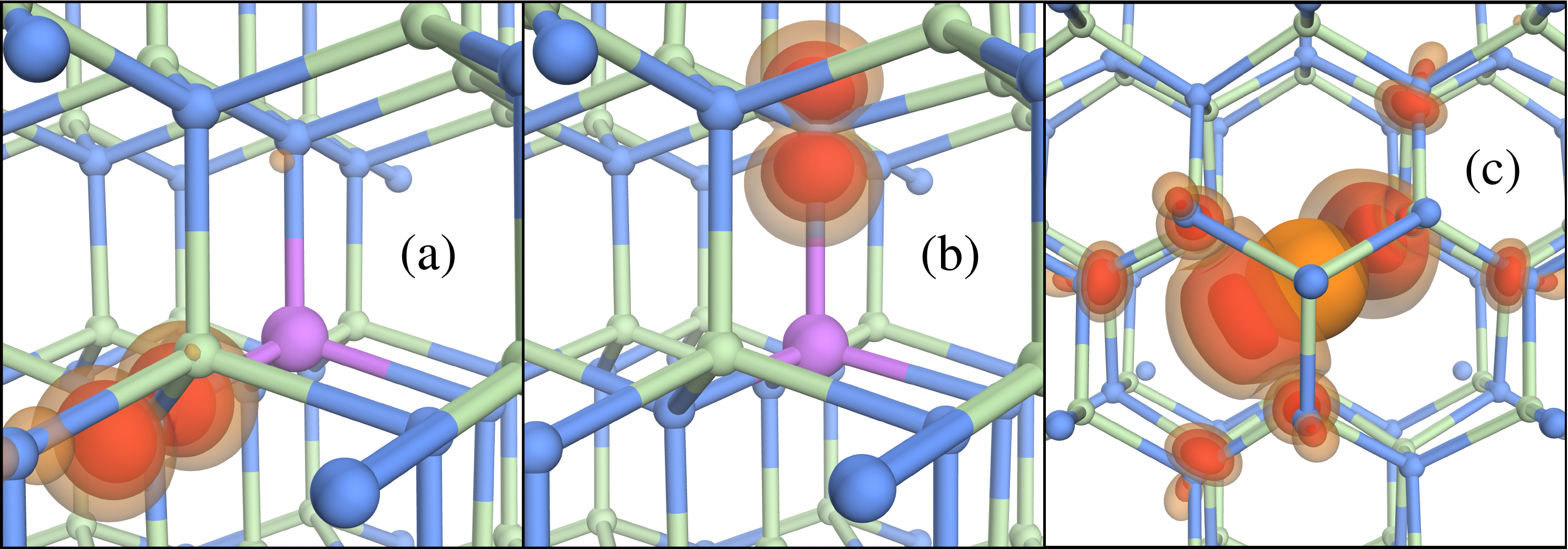}
\caption{(Color online) Calculated spin density resulting from (a) a
  Mg$_{\text{Ga}}^{0}$-associated hole localized on a neighboring N in
  the basal plane; (b) a Mg$_{\text{Ga}}^{0}$-associated hole
  localized on a neighboring axial N; and (c) a N vacancy. Light
  gray/green (darker gray/blue) spheres represent Ga (N) atoms. The
  darkest grey sphere represents a Mg atom in (a) and (b) (purple) and
  a vacancy in (c) (orange). Spin densities are indicated by (red)
  isosurfaces of density (au) 0.05, 0.025, and 0.01 for (a) and (b)
  and 0.01, 0.005, 0.0025 for (c).}
\label{spinden}
\end{figure}

We first consider the Mg$_{\text{Ga}}$ defect. We determine the
formation energy of defect $X$ ($E_f[X]$) as:
\begin{equation}
E_f[X] = \Delta E(X) - \sum_i n_i \mu_i + q E_F,
\label{formener}
\end{equation}
where $\Delta E(X)$ is the difference in energy between the embedded
cluster with and without $X$, $n_i$ is the number of atoms of species
$i$ added ($n_i > 0$) or subtracted ($n_i < 0$) in forming $X$,
$\mu_i$ is the chemical potential of species $i$, $q$ is the charge of
$X$, and $E_F$ is the Fermi energy. For $q=0$, there are two localised
hole configurations in the first coordination shell: either on a N in
the basal plane of the surrounding tetrahedron of N atoms or on the
axial N (for the related spin densities see Fig.~\ref{spinden} (a) and
(b)). Using Eq.~\ref{formener}, we find that, in anion-poor conditions
(see the Supplementary Information for details)
$E_f[Mg_{\text{Ga}}^0]$, with the hole localised on an axial N, is
1.928 eV, in excellent agreement with previous
calculations~\cite{gan_dft_mg_defect_holetrap_hybrid_prl_lyons2012}. In
anion-rich conditions, this energy changes to 0.783 eV. The axial hole
energy is more favourable (by 0.063 eV) than that of the configuration
with the hole localised on a N in the basal plane. If we consider
Mg$_3$N$_2$ as the source of Mg, rather than Mg metal, we obtain a
higher formation energy of 2.375 eV (2.756 eV) for anion-rich
(anion-poor) conditions.

The energy to dissociate a hole, according to the
reaction:
\begin{equation}
\text{Mg}_{\text{Ga}}^0 \rightarrow \text{Mg}_{\text{Ga}}^- + h^+
\label{mgtrap}
\end{equation} 
is highly unfavourable at 1.404 eV. This result differs from the low
value of 0.26 eV obtained using periodic supercell
models~\cite{gan_dft_mg_defect_holetrap_hybrid_prl_lyons2012}, which
may suffer from incomplete cancellation of the electron
self-interaction~\cite{sio2_al_local_clusters_prb_pacchioni2000,
  footnote1} and treatment of long-range polarisation, as well as an
absence of a well-defined reference~\cite{footnote2}. Our result,
however, is consistent with earlier work on thermodynamical doping
limits in GaN~\cite{doping_limits_chemmater_walsh2013}, which showed
that free holes are unstable with respect to point defect formation
(agreeing with the natively $n$-type nature of GaN). Indeed,
considering the compensation of holes by the formation of
V$_{\text{N}}$:
\begin{equation}
h^+ + \frac{1}{3}\text{N}_{\text{N}}^0 \leftrightarrow
\frac{1}{3}\text{V}_{\text{N}}^{3+} +
\frac{1}{6}\text{N}_2,
\label{holecomp}
\end{equation}
we determine a reaction energy of -1.245 eV, i.e. the balance is far
to the right, indicating the thermodymical instability of
holes. Consequently, doping with low levels of Mg will not result in
free-hole formation, instead compensation by 
V$_{\text{N}}$ will occur. We stress that our calculations apply to
the dilute limit, under the assumption of thermodynamical
equilibrium. For certain designs and/or synthetic procedures, where
kinetic effects may dominate, or at higher [Mg], where
the formation of complexes or phase segregated nanostructures may
occur, the balance of the reaction (Eq.~\ref{holecomp}) could
shift to the left.

\begin{figure}[ht!]
\centering
\vspace{0.2cm}
\includegraphics*[width=1.0\linewidth]{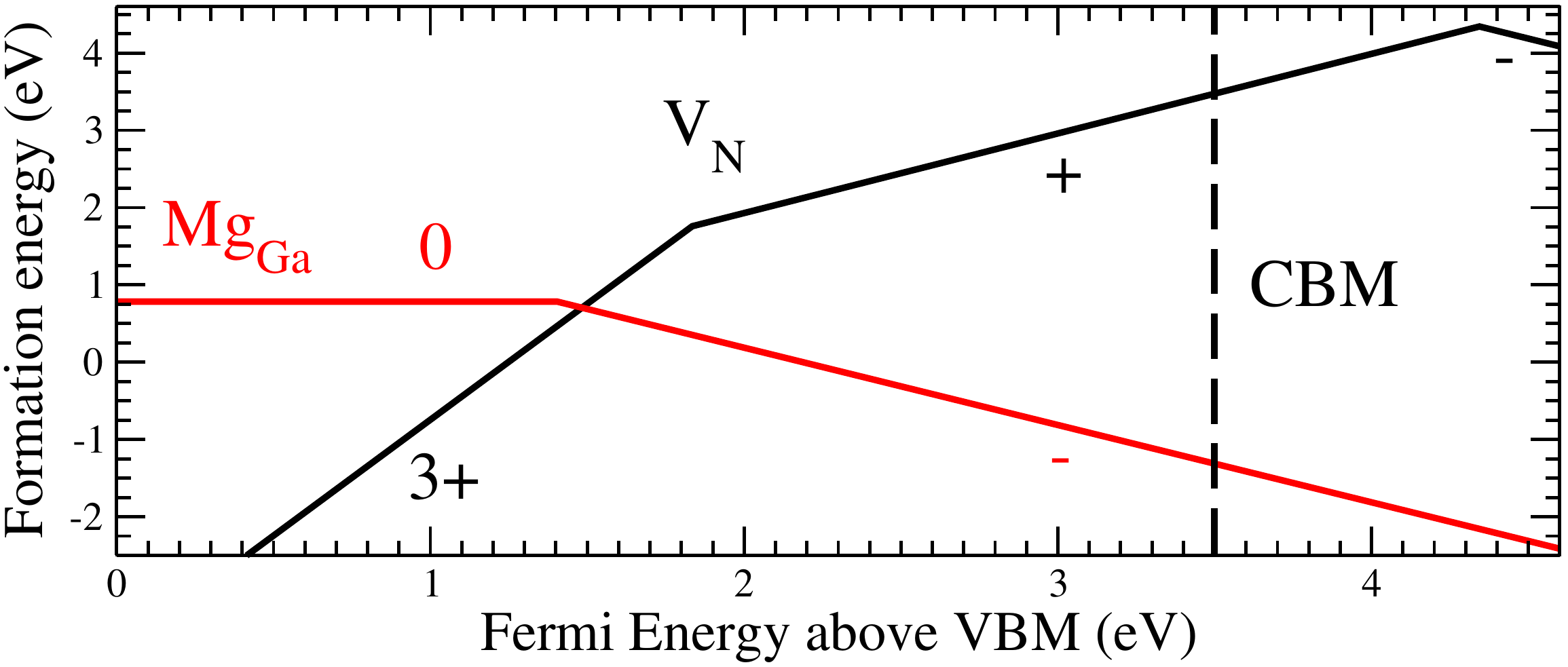}
\caption{(Color online) Formation energy of V$_{\text{N}}$ (black
  line) and Mg$_{\text{Ga}}$ (light gray/red line) as a function of
  Fermi energy above the valence band maximum (VBM). Anion-rich
  conditions are assumed. The position of the conduction band minimum
  (CBM) is indicated by the broken line. For each value of Fermi
  energy, only the most stable defect charge state is shown, with a
  change in slope indicating a change in charge state.}
\label{formener}
\end{figure}

We show the formation energy of V$_{\text{N}}$ (assuming N-rich
conditions), determined using Eq.~\ref{formener}, as a function of
$E_F$ relative to the VBM in Fig.~\ref{formener} (see
Fig.~\ref{spinden} (c) for the related spin density).  We find that,
in $n$-type conditions, the V$_{\text{N}}$ is singly ionised,
indicating that it is shallow. In $p$-type conditions the formation of
V$_{\text{N}}^{3+}$ becomes spontaneous, indicating the instability of
free holes.  V$_{\text{N}}^{2+}$ is unfavourable at any value of $E_F$
within the gap. V$_{\text{N}}^{+}$ and V$_{\text{N}}^{3+}$ are equally
favourable at $E_F = 1.835$ eV, a result significantly higher than
those determined using plane-wave basis set
calculations~\cite{GaN_review, gan_dft_vn_lum_mg-doped_apl_yan2012}
(see the above discussion on self-interaction
errors). V$_{\text{N}}^{-}$ becomes favorable in the regime of
degenerate $n$-type conduction at $\sim0.9$ eV above the CBM, where it
is expected to pin $E_F$. Integrating the density of states up to this
$E_F$ gives an $n$-type concentration of $\sim10^{20}$ cm$^{-3}$,
which has been observed in some undoped $n$-type
samples~\cite{iii_nitrides_manasreh}.

\begin{figure}[ht!]
\centering
\vspace{0.2cm}
\includegraphics*[width=1.0\linewidth]{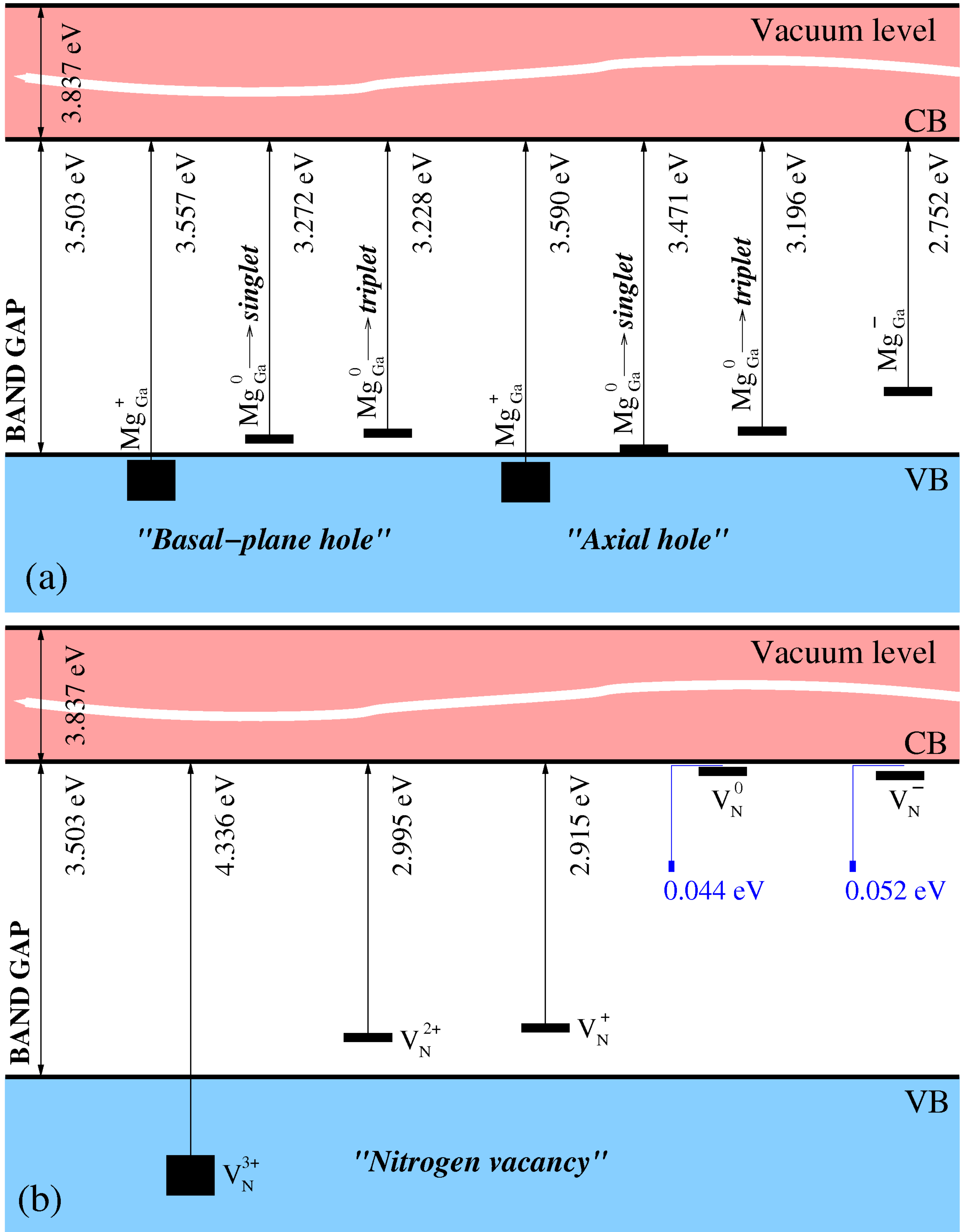}
\caption{(Color online) Calculated vertical ionisation levels of (a)
  Mg$_{\text{Ga}}$ and (b) V$_{\text{N}}$ in the relevant charge
  states, shown relative to the valence band (VB), conduction band
  (CB) and vacuum level. In (a) the two hole configuration cases,
  consisting of localisation on a N either in the basal plane or along
  the $c$ axis, are shown. For the neutral state, ionisation to either
  singlet or triplet states are included. The large black boxes
  indicate resonance states in the VB.}
\label{ion}
\end{figure}

In Fig.~\ref{ion} (a), we show the calculated vertical ionisation
energies (IEs) of Mg$_{\text{Ga}}$ relative to the conduction and
valence bands. The IE is defined as the difference between the energy
of a defect in charge state $q$, and the energy of the defect in the
same configuration, but with charge $q+1$. The VBM relative to the
vacuum level is determined by considering ionisation of an electron
from the bulk system. We include ionisation of Mg$_{\text{Ga}}^{0}$
and Mg$_{\text{Ga}}^{+}$ with both hole configurations (i.e. with the
hole localised on a basal plane N or axial N), as both should be
accessible from electron excitation experiments such as PL. For the
same reason, we include both possible final states, singlet and
triplet, for the case of ionisation of Mg$_{\text{Ga}}^{0}$. Such IEs
correspond to emission energies of photoexcited electrons recombining
with these defect levels, as would be observed in PL experiments,
where, after excitation, atoms typically do not have adequate time to
fully relax. Our calculations are in excellent agreement with the low
$T$ and [Mg] PL spectra observed
experimentally~\cite{gan_expt_mgdoped_pl_old+new_2lines_mg_jap_monemar2014,
  gan_expt_mgdoped_pl_290mev_level_apl_smith1996}. We reproduce well
the DAP PL peaks at 3.21 and 3.27 eV,~\cite{footnote3} and the ABE
peak at 3.466 eV. The 2.75 eV IE of Mg$_{\text{Ga}}^{-}$, although in
approximate agreement with the observed BL peak, is unlikely to be
observed due to its high formation energy (1.404 eV).

The vertical IEs of V$_{\text{N}}$ are, similar to the case of
Mg$_{\text{Ga}}$, shown in Fig.~\ref{ion} (b). We find that
V$_{\text{N}}^{0}$ is a shallow donor, having a vertical IE at 44 meV
below the CBM, in good agreement with measurements using TAS and
electron irradiation
techniques~\cite{n_vac_optic_admittance_evwaraye_jap_2014,
  gan_expt_eirradiate_donor_nvac_prl_look1997}. This level may also
account for the 3.46 eV PL peak observed in a wide range of
divalently-doped and undoped GaN
samples~\cite{gan_expt_mgdoped_pl_old+new_2lines_mg_jap_monemar2014,
  gan_expt_mgdoped_pl_290mev_level_apl_smith1996,
  gan_expt_pl_mgdoped_jap_reschikov2005}.  V$_{\text{N}}^{-}$ is also
stable and shallow with a vertical IE 52 meV below the CBM. The
stability and near degeneracy of the V$_{\text{N}}^{0}$ and
V$_{\text{N}}^{-}$ states facilitates Fermi level pinning near the
CBM. When the V$_{\text{N}}$ donates electrons and relaxes to its
equilibrium configurations, the resulting IEs are deep, and in
excellent agreement with the 2.95 eV BL peak that is observed at
higher [Mg] and at higher
$T$~\cite{gan_expt_mgdoped_pl_old+new_2lines_mg_jap_monemar2014,
  gan_expt_mgdoped_pl_290mev_level_apl_smith1996}. We associate these
levels with the BL as there will be more ionised V$_{\text{N}}$
present at higher $T$ and also at higher [Mg] due to an
increased number of compensating V$_{\text{N}}$.

\begin{table*}[ht!]
\caption{Calculated photoluminescence (PL) transitions, from total
  energy calculations, for different hole configurations and
  corresponding optical defect ionization levels compared with
  relevant experimental results taken from
  Refs.~\cite{gan_expt_mgdoped_pl_old+new_2lines_mg_jap_monemar2014,
    gan_expt_mgdoped_pl_290mev_level_apl_smith1996,
    gan_expt_pl_mgdoped_jap_reschikov2005,
    gan_mgdoped_224level_materressoc_johnson1997,
    gan_expt_zn_hg_li_applphys_ejder1974,
    gan_expt_zndopedI_jap_monemar1980,
    gan_expt_zndopedII_jap_monemar1980,
    gan_expt_bedoped_jap_ilegems1972, gan_expt_pl_jap_pankove1976,
    gan_be_semicondscitech_sanchez1998}.} \centering
\begin{ruledtabular}
\begin{tabular} { c | c c  c c c | c c  c }
& \multicolumn{5}{ c |}{Photoluminescence (eV)} & \multicolumn{3}{ c }{Levels (meV)} \\ 
& \multicolumn{2}{ c }{basal-plane hole} & \multicolumn{2}{ c }{axial hole} & Experiment & & & Experiment \\ \hline
Final state & singlet & triplet & singlet & triplet & & 0 & 1- & \\ \hline
Be & 3.410 & 3.427 & 3.454 & 3.456 & 3.384, 3.466 & 93, 76, 49, 47 & 624 & 90 \\ 
Mg & 3.272 & 3.228 & 3.471 & 3.196 & 3.21, 3.27, 3.466 & 275, 231, 307, 32 & 751 & 224, 290 \\ 
Zn & 3.144 & 3.201 & 3.068 & 3.195 & 3.100, 3.45 & 359, 302, 435, 308 & 975 & 340, 400, 480 \\ 
Cd & 2.845 & 2.929 & 2.814 & 2.934 & 2.8, 2.937, 3.455 & 658, 574, 689, 569 & 1197 & 560 \\ 
Hg & 2.584 & 2.666 & 2.587 & 2.694 & 2.70 & 919, 837, 916, 809 & 1455 & 410, 800 \\ 
\end{tabular}
\end{ruledtabular}
\label{others}
\end{table*}

In Table~\ref{others} we present our calculated IEs associated with
Be$_{\text{Ga}}$, Mg$_{\text{Ga}}$, Zn$_{\text{Ga}}$,
Cd$_{\text{Ga}}$, and Hg$_{\text{Ga}}$, along with the resulting
defect levels, compared with relevant experiment. In each case the
agreement is good. For all cases there is an observed 3.46 eV peak,
but only for the cases of Be$_{\text{Ga}}$ and Mg$_{\text{Ga}}$ can we
attribute it to an ABE. For the other dopants we attribute the peak to
the compensating V$_{\text{N}}$ (which may also play a role in Be and
Mg doping).

We therefore arrive at a simple explanation for what occurs when Mg or
other divalent metal dopants are added to GaN in small
concentrations. Isolated Mg$_{\text{Ga}}$ strongly trap holes and,
therefore, do not contribute to $p$-type conduction, instead hole
carriers will be compensated by the formation of V$_{\text{N}}$, a
result that follows for other divalent metal
dopants~\cite{footnote4}. The V$_{\text{N}}$ are shallow donors, with
the near degeneracy of the neutral and negative charge states pinning
the Fermi level close to the CBM, giving rise to the native $n$-type
conductivity. Our calculated IEs are in excellent agreement with the
relevant PL spectra. Furthermore, the V$_{\text{N}}$ level can give
rise to the 3.46 eV peak observed in a wide range of doped and undoped
samples, as it will be present as a compensating centre (for the case
of Mg and Be doping the 3.46 eV peak can also be attributed to ABEs,
which for Mg doping agrees with
experiment~\cite{gan_expt_mgdoped_epr_prb_glaser2002,
  gan_expt_pl_mgdoped_jap_reschikov2005}). A comprehensive description
of PL and conductivity phenomena in GaN lightly-doped with Mg, Be, Zn,
Cd, and Hg is thus provided, without the need to propose, in the
technologically significant case of Mg doping in particular,
clustering of Mg$_{\text{Ga}}$ and V$_{\text{N}}$, or H
impurities. The latter may, however, play an important role in
heavily-doped GaN, which is less well characterised or understood. The
key feature of such a material is the presence of inverted Mg-rich
pyramidal domains~\cite{gan_expt_mgdoped_pyramids_apl_vennegues2000},
which could trap interstitial N, making vacancy formation
unfavourable, and lead to lower hole ionisation potentials. Such
hypotheses, however, require apropriate investigation and are beyond
the scope of this study.

In summary, we have comprehensively studied defect formation associated
with divalent metal doping in GaN, using a multiscale approach. Our
results explain in detail the process, by which low levels of divalent
dopants are compensated by V$_{\text{N}}$, and are in excellent
agreement with available PL experimental data.

\section*{Acknowledgment}
The authors acknowledge funding from EPSRC grants ED/D504872,
EP/I01330X/1, EP/K016288/1. M. M. acknowledges
support from Accelerys Ltd. The authors also acknowledge the use of
the UCL Legion High Performance Computing Facility (Legion@UCL) and
associated support services, the IRIDIS cluster provided by the EPSRC
funded Centre for Innovation (EP/K000144/1 and EP/K000136/1), and the
ARCHER supercomputer through membership of the UK's HPC Materials
Chemistry Consortium (EPSRC grant EP/L000202). A. W. thanks C. G. Van
de Walle (UCSB) for useful discussions. A. W. and D. O. S. acknowledge
membership of the Materials Design Network. We would like to thank
C. Humphreys, T. D. Veal, and K. P. O'Donnell for useful discussions.

\begin{thebibliography}{10}
\bibitem{footnote3}
Here we are assuming a
  donor level of $30-50$ meV below the CBM, but excluding the Coulomb
  attraction energy between the donor and acceptor pair, as simply
  subtracting the donor level from the acceptor level values is
  equivalent to assuming infinite separation.
\end{thebibliography}



\end{document}